\documentstyle[12pt,twoside,fleqn,espcrc1,epsfig]{article}

\newcommand{\AmS}{{\protect\the\textfont2
  A\kern-.1667em\lower.5ex\hbox{M}\kern-.125emS}}

\title{Random Matrix Theory and QCD at nonzero chemical Potential}

\author{J.J.M. Verbaarschot\address{Department of Physics and Astronomy,\\
                                    University at Stony Brook,\\
                                    Stony Brook, NY\,11794, USA} }

\begin{document}
\maketitle

\begin{abstract}
In this lecture we give a brief review of chiral Random Matrix Theory (chRMT)
and its applications to QCD at nonzero chemical potential. 
We present both analytical arguments involving chiral perturbation
theory and numerical evidence from lattice QCD simulations
showing that correlations of the smallest Dirac eigenvalues are described by
chRMT.  We discuss the range of validity of chRMT and emphasize 
the importance of universality. For chRMT's at $\mu \ne 0 $ we 
identify universal properties of the Dirac eigenvalues and study   
the effect of quenching on the distribution of Yang-Lee zeros.
\end{abstract}

\newcommand{\be}{\begin{eqnarray}}
\newcommand{\ee}{\end{eqnarray}}

\section{INTRODUCTION}

There are strong indications from hadronic phenomenology 
that chiral symmetry is 
broken spontaneously at low temperatures. 
However, mainly through lattice QCD simulations,
it has become generally accepted 
that chiral symmetry is restored above a critical temperature of about
$140\, MeV$  (see reviews by DeTar {\cite{DeTar}}, Ukawa {\cite{Ukawa}},
Smilga {\cite{Smilref}} and Karsch \cite{Karsch}
for recent results on this topic). 
The situation at nonzero chemical potential is much less clear. Because
of the phase of the fermion determinant it is not possible to perform
Monte-Carlo simulations, and it seems that one has to give up 
calculations based on first principles. 
It is certainly not possible to make progress by a 
gradual improvement of existing techniques.
Instead, radically different approaches have to be developed.

The main source of the problem is the loss of Hermiticity  at
nonzero chemical potential. In this lecture we study this problem by
means of a much simpler 
chiral Random Matrix Theory (chRMT) 
with the global symmetries of the Dirac operator
\cite{SVR,JV,Tilo,Stephanov}.
This model has the remarkable property that, in spite of the fact that
it can be solved analytically, it cannot be solved numerically. 
In the broken phase, numerical convergence is only obtained for an 
exponentially large ensemble \cite{HJV}. 
If one is not able to develop an algorithm for 
this simple model, progress in QCD will be elusive.

Originally, chRMT was introduced to describe the correlations of
QCD Dirac eigenvalues. Our interest in the Dirac spectrum is based on the 
relation between the chiral condensate, $\Sigma$,
and the spectral density per unit of space time volume, $V$, given by
$\Sigma = \lim {\pi \rho(0)}/V$  \cite{BC}.
Note that the thermodynamic limit has to be taken before the chiral limit.
The spectral density is defined by 
$\rho(\lambda) = \sum_k \delta(\lambda- \lambda_k)$, 
where the $\lambda_k$ are the eigenvalues of the Dirac operator.
Because of this relation the eigenvalues near zero are spaced as 
$1/\rho(0) = \pi/\Sigma V$. In order to study the approach to the thermodynamic
limit it is natural to introduce the microscopic limit in which $u=\lambda V 
\Sigma$ is kept fixed for $V\rightarrow \infty$, and the microscopic spectral
density \cite{SVR}
\be
\rho_S(u) = \lim_{V\rightarrow \infty} \frac 1{V\Sigma} \langle
\rho(\frac u{V\Sigma})\rangle.
\label{rhosu}
\ee
Our claim is that the microscopic spectral density of the QCD Dirac operator
is given by chRMT. This has been confirmed both by analytical arguments
using partially quenched chiral perturbation theory \cite{Toublan}, and
lattice QCD \cite{Tiloprl,Ma,many,Guhrth,Tilomass} 
and instanton liquid  \cite{Vinst,Osborn} simulations.
In addition to this, correlations in the bulk of the spectrum are given
by chRMT \cite{HV,HKV,Markum,Guhrth} as well. Random matrix theory
can describe only those observables which are universal, i.e., which
are stable against large deformations of the random matrix ensemble.
By now it has been well established that both the microscopic spectral density 
and the spectral correlations are strongly universal 
\cite{Damgaard,Brezin,GWu,Sener1,andystudent,Tilodam,Senerprl,Widom},
\cite{Dampart}.
Before discussing the random matrix model at nonzero chemical potential,
we wish to establish the domain of applicability of chRMT  
at $\mu = 0$. Starting from partially quenched Chiral 
Perturbation Theory \cite{pqChPT} we will argue that, below 
a scale of $\Lambda_{QCD}/\sqrt V$, eigenvalue correlations 
are given by chRMT. 
This scale is the equivalent of the Thouless
energy in mesoscopic physics (see 
\cite{HDgang,Beenreview,Montambaux} for recent reviews). 
The existence of such a scale has been confirmed by
lattice QCD \cite{many,Guhrth} and instanton liquid \cite{Osborn} 
simulations. 
The interpretation is that the classical motion of the
quarks in the background gauge fields is chaotic.

Before proceeding to the main body  of this talk, 
let me stress that there are two different 
types of applications
of RMT. First, as the simplest model of a universality class,
and second, as a schematic model for a complex system. Typical  examples
in the first class are related to eigenvalue correlations \cite{HDgang}.
Perhaps, the most famous
example in the second class is the Anderson model for 
localization \cite{Anderson}.
In this lecture we will discuss a schematic
RMT model for the chiral phase transition at nonzero chemical potential.

\section{CHIRAL RANDOM MATRIX THEORY}

In this section we will introduce an instanton 
liquid {\cite{shurrev,diakonov}} inspired 
chiral RMT for the QCD partition function. 
With only its global symmetries as input  but otherwise Gaussian distributed 
matrix elements the chRMT for $N_f$ flavors and topological charge $\nu$
is defined by \cite{SVR,V}
\be
Z_{N_f,\nu}^\beta(m_1,\cdots, m_{N_f}) = 
\int DW \prod_{f= 1}^{N_f} \det({\rm \cal D} +m_f)
e^{-\frac{N\Sigma^2 \beta}4 {\rm Tr}W^\dagger W},
\label{zrandom}
\ee
where
\be
{\cal D} = \left (\begin{array}{cc} 0 & iW\\
iW^\dagger & 0 \end{array} \right ),
\label{diracop}
\ee
and $W$ is a $n\times m$ matrix with $\nu = |n-m|$ and
$N= n+m$. We interpret $N$ as the (dimensionless) volume of space
time. The matrix elements of $W$ are either real ($\beta = 1$, chiral
Gaussian Orthogonal Ensemble (chGOE)), complex
($\beta = 2$, chiral Gaussian Unitary Ensemble (chGUE)),
or quaternion real ($\beta = 4$, chiral Gaussian Symplectic Ensemble (chGSE)).
As is the case in QCD, we assume that $\nu$ does not exceed $\sqrt N$, so that,
to a good approximation, $n = N/2$. The Dyson index $\beta$ of a physical 
system is determined by its anti-unitary symmetries. If the square of
the anti-unitary symmetry operator is equal to the identity, we have 
$\beta = 1$. If its square is equal to minus the identity, we have $\beta = 4$.
If there no anti-unitary symmetries, the value of $\beta = 2$.

In this model chiral symmetry is broken spontaneously with chiral condensate
given by  $\Sigma = \lim_{N\rightarrow \infty} {\pi \rho(0)}/N$.
For complex matrix elements ($\beta =2$), which is appropriate for QCD 
with three or more colors and fundamental fermions, the
symmetry breaking pattern is $SU(N_f) \times SU(N_f)/SU(N_f)$
{\cite{SVR,SmV}} .

The average spectral density that follows from (\ref{zrandom}) 
has the familiar semi-circular shape.
The microscopic spectral density can be derived from the limit
(\ref{rhosu}) of the exact spectral density for finite $N$.
For $N_c = 3$, $N_f$ flavors and topological charge $\nu$
it is given by {\cite{VZ,V}}
\be
\rho_S(z) = \frac z2 \left ( J^2_{a}(z) -
J_{a+1}(z)J_{a-1}(z)\right),
\label{micro2}
\ee
where $a = N_f + |\nu|$. 
The spectral correlations in the bulk of the spectrum are given 
by the invariant random matrix ensemble with the same value
of $\beta$ {\cite{Kahn,Nagao}}. 

The description of an observable in terms of chRMT is based on universality,
i.e. on the stability against deformations of the
random matrix ensemble. At this moment I only mention the work of 
Akemann, Damgaard, Magnea and Nishigaki \cite{Damgaard}
who showed that the same microscopic
spectral density (\ref{micro2}) 
is obtained for an invariant probability distribution defined
by an arbitrary  polynomial potential.
The essence of their proof
is a remarkable generalization of the identity for
the Laguerre polynomials,
$\lim_{n \rightarrow \infty}  L_n( x/ n) =
 J_0(2 \sqrt x) $,
to orthogonal polynomials determined by the probability distribution.

\section{DOMAIN OF VALIDITY}

The domain of validity of chRMT is best discussed within the context of 
partially quenched chiral perturbation theory \cite{pqChPT,Toublan}. 
This is an effective field
theory for the low-energy limit of a QCD like theory which,
in addition to the usual $N_f$ sea quarks, contains 
$k$ valence quarks with mass $m_v$ and their super-symmetric
partners with the same mass.
In this framework it
is possible to calculate the valence quark mass dependence of the chiral 
condensate which is defined as
\be
\Sigma_v(m_v) = \frac 1N \int d\lambda \langle\rho(\lambda)\rangle
\frac{2m_v}{\lambda^2 +m^2_v}.
\label{sigmam}
\ee
The spectral density follows from the discontinuity of
$\Sigma_v(m_v)$,
\be
\frac{ 2\pi}N \langle \rho(\lambda)\rangle =
\left .{\rm Disc}\right |_{m_v = i\lambda}\Sigma(m_v)
\equiv \lim_{\epsilon \rightarrow 0}
\Sigma(i\lambda+\epsilon) - \Sigma(i\lambda-\epsilon) = \frac{2\pi}N \sum_k
\langle \delta(\lambda +\lambda_k)\rangle
\label{spectdisc},
\ee
where the average $\langle \cdots \rangle$ is with respect to the distribution
of the eigenvalues.
Similarly, the two-point spectral correlation function, given by
\be
\langle \rho(\lambda) \rho(\lambda') \rangle
= \frac 1{4\pi^2}\left . {\rm Disc}
\right |_{m_v = i\lambda, m_{v'}=i\lambda'}
  \sum_{k,l}\left\langle
 \frac 1{i\lambda_k +m_v}\frac 1{i\lambda_l +m_{v'}}\right\rangle,
\ee
is related to the scalar susceptibility. Both the valence quark mass dependence
of the chiral condensate and the scalar susceptibility can be calculated 
from the pqChPT partition function. 
In both cases we can identify an important scale where the mass
of the Goldstone modes containing a valence quark is equal to
the inverse length of the box \cite{GL,LS}. 
Using the relation $M= (m + m')\Sigma/F^2$, 
where $F$ is
the pion decay constant, we find from $ML =1$ that, in terms of the valence
quark mass $m_v$, this scale is given by \cite{Osborn,Janik}
\be
E_c = \frac{F^2}{\Sigma L^2}.
\ee
For $m_v \ll E_c$ we have shown that the valence quark mass dependence and
the scalar susceptibility  can be obtained from the zero-momentum component
of the pqChPT partition function and are given by the result for chRMT. 
The conclusion is that, if
pqChPT describes correctly the low energy limit of QCD, we have shown that
the correlations of the Dirac eigenvalues close to zero are given by 
chRMT.

Such a picture is well-known from mesoscopic physics. In this context 
$E_c$ is defined as the inverse tunneling time of an electron through
the sample which is given by $E_c = {\hbar D}/{L^2}$, where $D$ 
is the diffusion constant. Another scale that enters in 
these systems is the elastic scattering time $\tau_e$. Based on these scales
one can distinguish three different regimes
for the energy difference, $\delta E$, that enters in the two-point correlation
function \cite{Altshuler}: i) the ergodic regime
for $\delta E \ll E_c$, ii) the diffusive domain for $E_c \ll \delta E \ll 
\hbar/\tau_e$  
and iii) the ballistic regime for $\delta E \gg \hbar/\tau_e$.
On time scales corresponding to the ergodic regime an initially localized
wave packet covers all of phase space. In this domain the eigenvalue
correlations are given by RMT. On time scales corresponding to 
the diffusive domain an initially localized
wave packet explores only part of the phase space resulting 
in eigenstates  with wavefunctions
that are localized in different regions. The corresponding eigenvalues
are show weaker correlations which are no longer given by RMT.
For earlier applications of localization theory to the chiral phase transition
we refer to \cite{shuryak}. 

Based on these ideas we can interpret the Dirac spectrum as the energy levels
of a system in four  Euclidean dimensions and one artificial time dimension.
According to the Bohigas conjecture \cite{Bohigas}
the eigenvalue correlations are given by
RMT if and only if the corresponding classical motion is chaotic. We thus
conclude  that the classical time evolution of quarks in the Yang-Mills
gauge fields is chaotic.


These ideas have been tested by means of lattice QCD \cite{many,Guhrth} 
and instanton liquid \cite{Osborn} simulations. It has been found
that the eigenvalue correlations
are given by chRMT up the predicted scale of  $ E_c= F^2/\Sigma L^2$.


\section{LATTICE QCD RESULTS}

In this section we consider correlations of lattice QCD Dirac eigenvalues.
For a Dirac operator with a $U_A(1)$ symmetry the eigenvalues 
occur in pairs $\pm \lambda_k$. Therefore  we have to distinguish
two different regions: the region near zero virtuality and the bulk of the
spectrum. The $U_A(1)$ symmetry is absent for the 
Hermitean Wilson Dirac operator.

The class of random matrix ensembles 
is determined by anti-unitary symmetries.
In the case of an $SU(2)$ color group,
the anti-unitary symmetries of the Kogut-Susskind (KS)
and the Wilson Dirac operator are given by {\cite{Teper,HV}},
\be
[D^{KS}, \tau_2 K] = 0, \quad {\rm and} \quad 
[\gamma_5 D^W, \gamma_5 CK\tau_2] = 0.
\ee
Because $(\tau_2 K)^2 = -1$ and $(\gamma_5 CK\tau_2)^2 =1$,
we find a Dyson index $\beta = 1$ for Wilson fermions and $\beta = 4$ for
KS fermions. The KS Dirac operator for two colors is thus described by the 
chGSE. In the absence of a $U_A(1)$  symmetry, the Wilson Dirac operator for 
two colors is described by the GOE \cite{HV}.
For three or more colors there are no anti-unitary symmetries
and the relevant ensembles are the chGUE and GUE for KS and Wilson fermions,
respectively.

In order to separate the correlations from the average spectral density,
the eigenvalues are rescaled according to the  local average level spacing. 
Under the assumption of spectral ergodicity, eigenvalue correlations 
in the bulk of the spectrum  can be calculated by spectral averaging
instead of ensemble averaging. For subtle differences between the two
we refer to \cite{Guhrth}.
Eigenvalues calculated by Kalkreuter {\cite{Kalkreuter}} 
for the $N_c = 2$ KS and Wilson Dirac operators  
for lattices as large as $12^4$ have been analyzed
by means of spectral averaging \cite{HV,HKV}.
For different values of $\beta$, ranging from strong coupling to weak coupling,
it was found that correlations are in complete agreement with  chRMT 
for distances as large as 100 level spacings.
Recently, correlations in the bulk of the spectrum were also investigated
for $N_c = 3$ \cite{Markum,markumbiel}. Agreement with chRMT
 was even found for values of $\beta$ well above the 
deconfinement transition. For a discussion of global Wilson spectra we refer
to \cite{Jurkiewicz,Hehl}.


Spectral ergodicity cannot
be exploited in the study of the microscopic spectral density.
In order to gather sufficient statistics 
 one has to generate 
a large number of independent QCD Dirac spectra.
\begin{figure}[!ht]
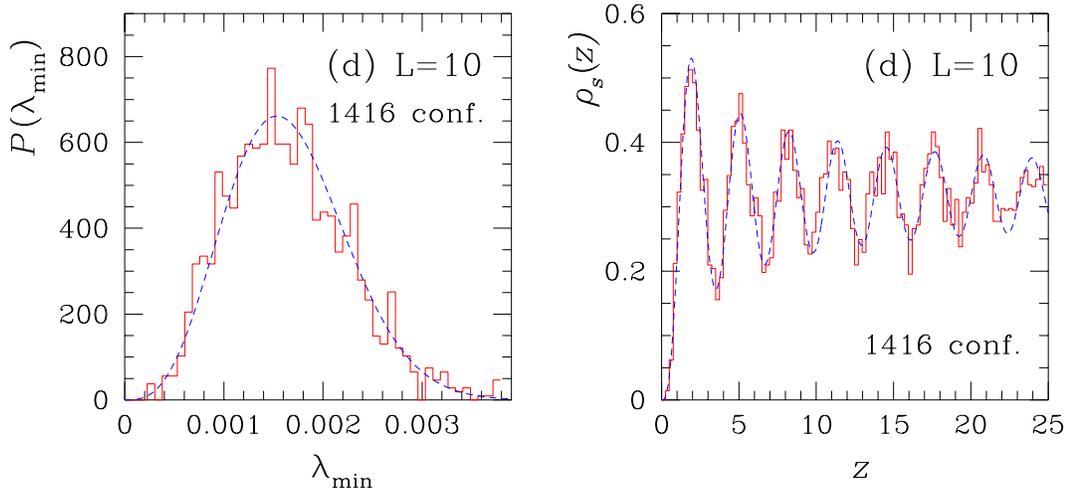

\vspace*{-2cm}
{\makebox{\epsfig{file=tilo10small.ps,
width=70mm,angle=0}}}
{\makebox{\epsfig{file=tilo10micro.ps,
width=70mm,angle=0}}}
\vspace*{-2.75cm}
\caption{The distribution of the smallest eigenvalue (left) and the
microscopic spectral density (right) of the Kogut-Susskind Dirac operator
for two colors and $\beta = 2.0$. }
\label{fig4}
\vspace*{-0.5cm}
\end{figure}
The analysis of these spectra  convincingly shows 
that the smallest eigenvalues
are correlated according to chRMT \cite{Tiloprl,Ma,many,Tilomass,Guhrth}.
Results for 1416 configurations \cite{Tiloprl}
on  $10^4$ lattice for $\beta = 2$ are shown in Fig. (\ref{fig4}).
For more detailed results,
including results for the two-point correlation function,
we refer to the original work. In Fig. 4  we show the distribution of 
the smallest eigenvalue (left) and the microscopic spectral density (right).
The lattice results are given by the full line. The dashed curves represent
the random matrix results. 
We emphasize that the theoretical curves have been
obtained without any fitting of parameters. The input parameter, the
chiral condensate, is derived from the same lattice calculations. 
Recently, the same analysis {\cite{Tiloprl}} 
was performed for  $\beta = 2.2$
and for $\beta =2.5$ on a $16^4$ lattice. In both cases
agreement with the random matrix predictions was found {\cite{Tiloprl}}.

It is an interesting question of how spectral correlations of KS fermions
evolve in the approach to the continuum limit. Certainly, the 
Kramers degeneracy of the eigenvalues remains. However, since Kogut-Susskind
fermions represent 4 degenerate flavors in the continuum limit, 
the Dirac eigenvalues should obtain an additional two-fold degeneracy.
We are looking forward to more work in this direction.

\section{APPLICATIONS OF chRMT AT $\mu \ne  0$}

In the continuum formulation of QCD the chemical potential enters in the
QCD partition function by the addition of the term $\mu \gamma_0$ to the
anti-Hermitean Dirac operator, i.e. ${\cal D} \rightarrow {\cal D} + \mu 
\gamma_0$. In a suitable chiral
basis, in which the matrix elements of $\langle 
\phi_R^k|\gamma_0|\phi_L^k\rangle= \delta_{kl}$, the corresponding modification
in the random matrix partition function  (\ref{zrandom})  
is to replace the Dirac matrix ${\cal D}$ by \cite{Stephanov}
\be
{\cal D}(\mu) =
\left ( \begin{array}{cc} 0 & iW + \mu\\
iW^\dagger +\mu & 0 \end{array} \right) \ .
\label{Diracmatter}
\ee

As is the case in QCD, a nonzero chemical potential violates the 
anti-Hermiticity of the Dirac operator and its eigenvalues are scattered in the
complex plane. For three or more colors 
this results in a complex fermion
determinant. The presence of this phase makes it
virtually impossible to perform Monte-Carlo simulations for the full theory.
The easy way out would be to ignore the fermion determinant altogether.
However, it was shown that the quenched approximation fails 
dramatically \cite{everybody,Barbour,Klepfish}. The 
 critical chemical potential is determined by the
pion mass instead of the nucleon mass. After some earlier work by 
\cite{Gibbs1,Gocksch} 
this puzzle was resolved convincingly by Stephanov \cite{Stephanov}
within
the context of chRMT at $\mu \ne 0$. He could show analytically that the
quenched approximation is not the $N_f \rightarrow 0$ limit of the full
theory but rather the $N_f \rightarrow 0$ limit of a theory in which the
fermion determinant is replaced by its absolute value. This theory can
be interpreted as a theory that  contains an equal number of quarks
and conjugate quarks. It was shown that in this theory chiral symmetry is 
broken spontaneously by the formation of a quark 
conjugate anti-quark condensate.
The critical chemical potential is then determined
by the mass of the corresponding Goldstone bosons with a net baryon number. 

Because the term $\mu \gamma_0$ does not change the anti-unitary 
symmetries of the Dirac operator  the 
classification for $\mu = 0$ is also valid at $\mu\ne 0$.
For example, the Kogut-Susskind Dirac operators 
for $N_c \ge 3$ and $N_c =2$ corresponds to the classes 
$\beta =2$ and $\beta = 4$, respectively, whereas 
the continuum Dirac operator for $N_c = 2$ corresponds to $\beta = 1$.

The fermion determinant is real both for $\beta= 1$ and $\beta = 4$.
This is obvious for $\beta = 1$. For $\beta = 4$ the reality follows from the
identity $q^* = \sigma_2 q \sigma_2$ for a quaternion real element $q$, and
the invariance of a determinant under transposition.
We thus conclude that Monte-carlo simulations can be performed for $\beta =1$
and $\beta =4$. For these values of $\beta$ the partition function can
also be interpreted in terms of quarks and conjugate quarks
and chiral symmetry will be  restored for arbitrarily,
small nonzero $\mu$, whereas a condensate of
a quark and a conjugate anti-quark develops. Indeed,
this phenomenon has been observed
in the strong coupling limit of lattice QCD
with two colors {\cite{Elbio}}.

Alternatively, one may write $\det {\cal D}+m=
\det \gamma_0({\cal D}+m)$ with the
fermion matrix \cite{Gibbs}
\be
\gamma_0({ \cal D}(\mu) + m)=
\left ( \begin{array}{cc} iW + \mu &m\\
m& iW^\dagger +\mu \end{array} \right). \ 
\ee
Let us study this model for $m = 0$. 
In terms of the eigenvalues, $\lambda_k$, of $W$
the baryon number density is simply given by
\be
n_q = \frac 1N \sum_k \left [\frac 1{\lambda_k + \mu}+
\frac 1{\lambda_k^* + \mu} \right ].
\ee
A natural electrostatic interpretation of this result is that $n_q$
is the electric field at location $\mu$ due to charges located at the
positions of the eigenvalues.
As was shown by Ginibre \cite{Gin}, the eigenvalues
of $W$ are distributed homogeneously in a disk in the complex
plane. For small $\mu$ we thus have that $n_q \sim \mu$  
in the quenched approximation.

For unquenched QCD  we expect  that $n_q = 0$ for 
$\mu < \mu_c\ne 0$.
By the electrostatic
analog this is only possible if there are no
eigenvalues in the domain $\mu < \mu_c$. The effect of the fermion determinant 
is to average the spectral density of $W$ to zero in this region. In the
random matrix model the phase of the fermion determinant does not 
result in a zero baryon density below $\mu_c$. Instead, $n_q$ becomes negative.
The reason is that chRMT  is a schematic model for the chiral 
phase transition in which the low temperature behavior has not been
implemented correctly. The claim is that taking into account all
Matsubara frequencies will restore this deficiency \cite{JaNo97}. 

This RMT partition function at $\mu \ne 0$ has inspired a great deal of 
interest. For recent literature on this topic we refer to
\cite{Feinberg-Zee,Nowakrecent,JaNo97,HJV}.

\subsection{Yang-Lee Zeros of the Partition Function}

In this section we discuss the distribution of Yang-Lee zeros 
\cite{Frank} in the
complex $\mu$ plane. In particular, we compare the zeros 
of the unquenched partition function for $N_f = 2$ and of the 
partition function with one flavor and one conjugate flavor
(which we will denote by $N_f = 1+1^*$). A recent controversial question 
raised in \cite{spanish} is
whether the phase quenched partition function with $N_f = 1+1^*$ 
can teach us something about the full theory. Zeros of polynomials have
been studied elaborately in the mathematics literature. At this moment
we only mention a limit theorem
for the distribution of zeros of polynomials with random coefficients
which states that it converges to the complex unit circle (see for example
\cite{randompol}). 

Since the chRMT partition function is a polynomial in $\mu$ it can be
factorized as 
\be
Z= \prod_k(\mu - \mu_k),
\ee
where the $\mu_k$ are the complex zeros of the partition function.
The baryon number density is thus given by \cite{Vink,Barbourqed}
\be
n_q = \frac 1N \sum_k \frac 1{\mu -\mu_k},
\ee
and can be interpreted as the electric field at the position $\mu$ from
charges located at $\mu_k$ in two dimensional complex $\mu$-plane. 

In QCD at $T= 0$, we expect
that $n_q = 0$ for $\mu < \mu_c$ which is only possible if, 
in the thermodynamic limit,  the number
of zeros with absolute value less than $\mu_c$ become negligible with
respect to the total number of zeros. 
At $\mu_c$ we expect a first order phase transition with a discontinuity
in the baryon number density. In terms of the zeros, this implies that 
they should coalesce into to a cut for $N\rightarrow \infty$. In order to have
$n_q = 0$ for $|\mu| < \mu_c$, this cut has necessarily to be a 
circle of radius $\mu_c$. Since the low-temperature limit has not been
implemented correctly in our schematic model we will find a different curve
for the location of the zeros.

\begin{figure}[!ht]
{\makebox{\epsfig{file=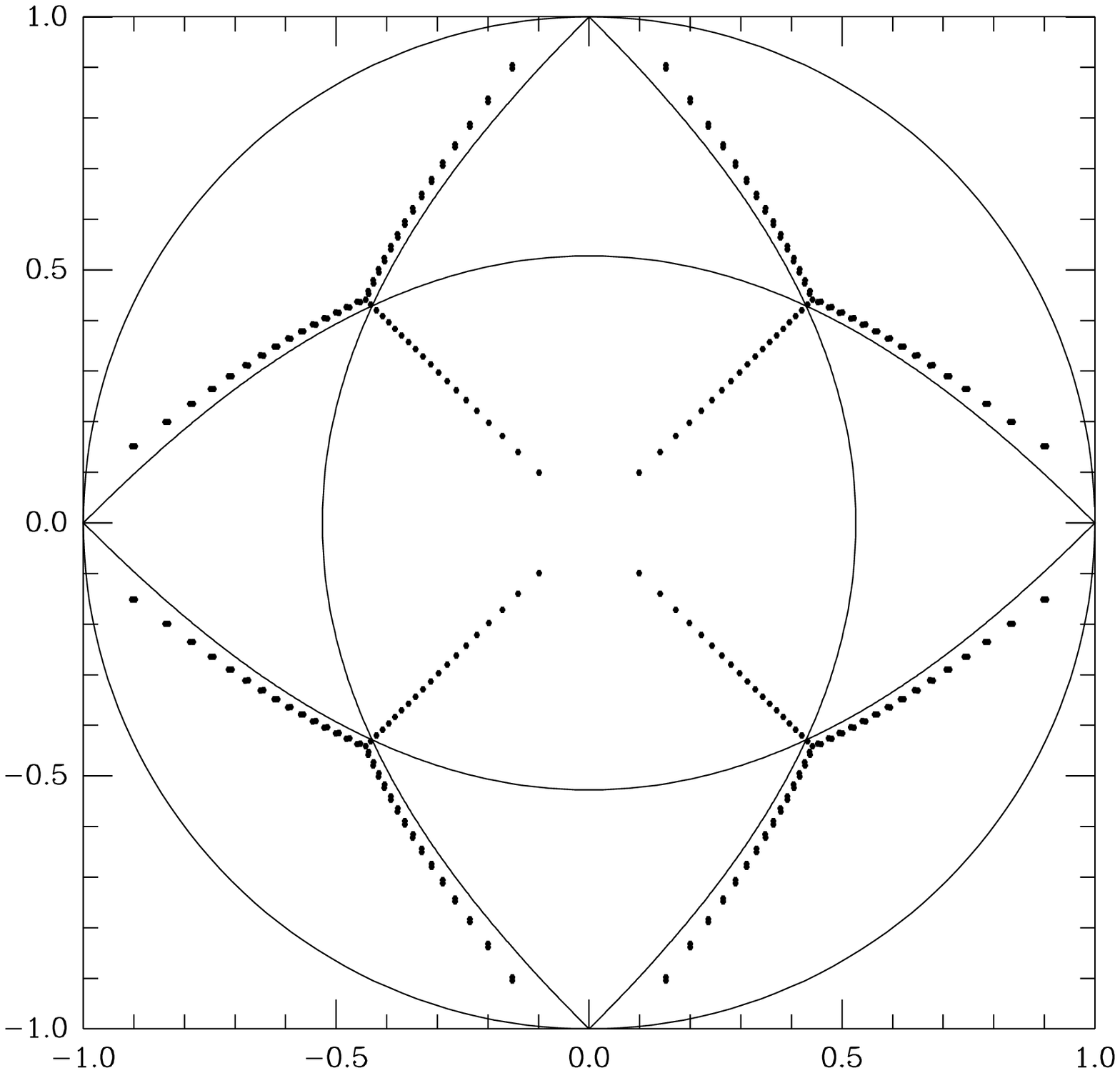,
width=75mm,angle=0}}}
{\makebox{\epsfig{file=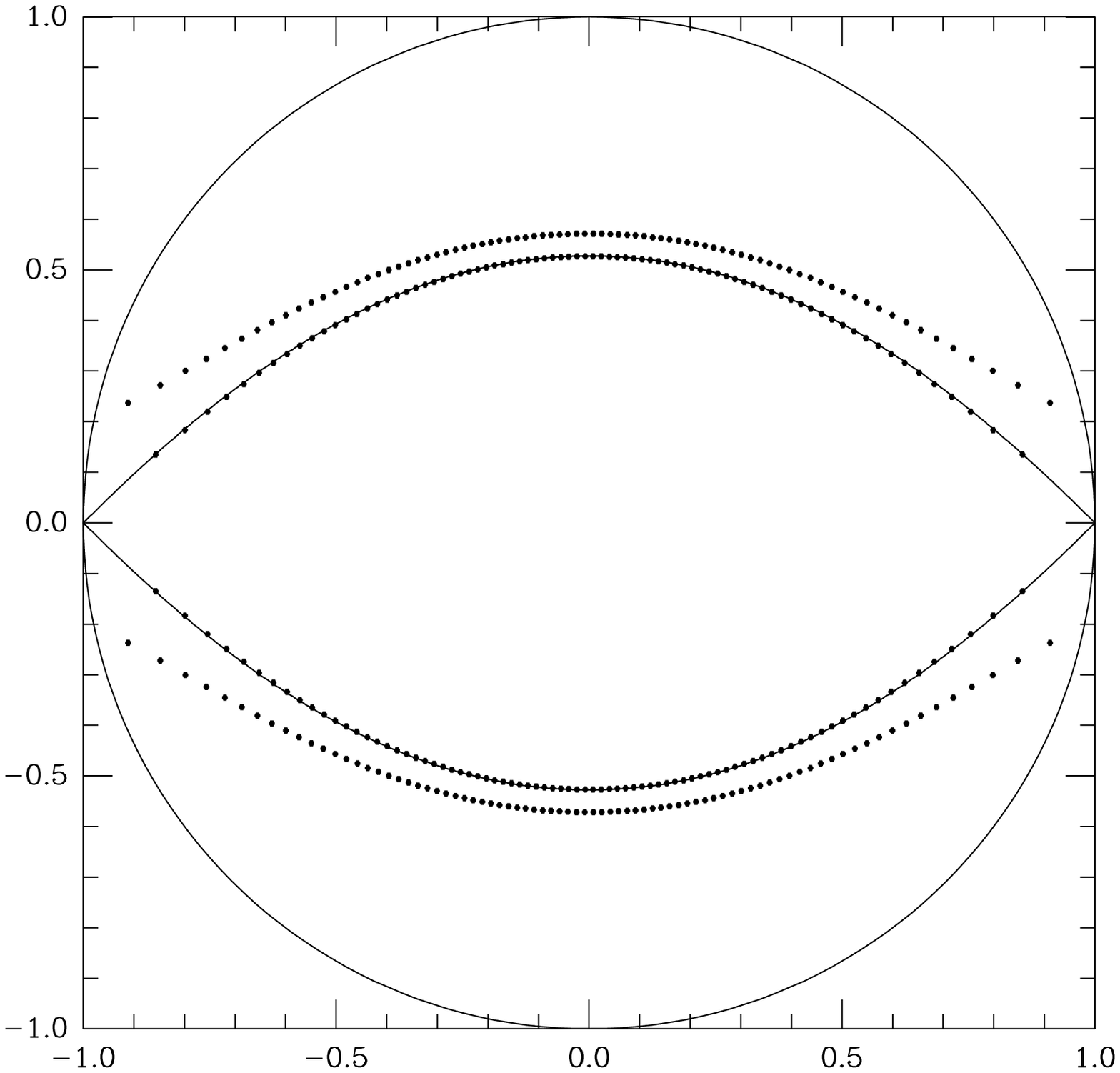,
width=75mm,angle=0}}}	
\vspace*{-1.0cm}
\caption{Yang-Lee zeros of the RMT partition function with $N_f = 1+1^*$
(left) and $N_f = 2$ (right). }
\label{Fignato}
\vspace*{-0.5cm}
\end{figure}

In terms of the  $\sigma$-representation 
the random matrix partition function for $N_f$ flavors can be rewritten 
\be
Z^{N_f}(m=0, \mu) = \int {\cal D} \sigma \exp [-{n\Sigma^2} {\rm Tr}
\sigma \sigma^\dagger] {\det}^{n } (\sigma\sigma^\dagger-\mu^2)
\ee
where $\sigma$ is an arbitrary complex $N_f\times N_f$ matrix. 
Since the value of the chiral condensate is given by the expectation value 
of $\sigma$, the
logarithm of the integrand can be interpreted as a Landau-Ginzburg functional
for the order parameter. From a saddle-point analysis it follows that this
partition function describes a first order phase transition
\cite{Stephanov}. Since the
equality of the real part of the free energies on opposite sides of 
the phase boundary
imposes one condition on the value of $\mu$ in the complex plane, the
phase boundary is given by a curve in the complex plane
\cite{Shrock}. In the thermodynamic
limit the zeros that are located on this curve should join into a cut.  
The result for this curve given by
\cite{Stephanov,HJV}
\be
{\rm Re}[ \mu^2 + \log(\mu^2) ] = -1
\label{mucurve}
\ee
is represented by the solid curve within the complex unit circle in
Fig. (\ref{Fignato}). It does not depend on the number of flavors 
for $n \rightarrow \infty$.

The $\sigma$-model representation of the partition function for one flavor
and one conjugate flavor is given by
\be
Z^{1+1^*}(m=0, \mu) = \int {\cal D} \sigma \exp [-{n\Sigma^2} {\rm Tr}
\sigma \sigma^\dagger] {\det}^{n } (\sigma k \sigma^\dagger k   
-\mu^2),
\ee
where $\sigma$ is an arbitrary complex $2\times 2$ matrix and
$k\equiv \sigma_3$.
This partition function is an exact rewriting of the partition function
(\ref{zrandom}) for $N_f =2$ with (\ref{Diracmatter})
but with the determinant replaced by its
absolute value. 
For even $n$ this partition function is invariant under $\sigma \rightarrow
\sigma \sigma_1$ and $\mu \rightarrow i\mu$. As a consequence it can
be factorized as a polynomial in $\mu^4$.   

The numerical results for the zeros of the partition function for
$N_f = 2$ (left) and $N_f = 1+1^*$ (right) 
are depicted by the full circles in Fig. (\ref{Fignato}). 
In both cases the total
number of zeros is equal to 320.
They  were
obtained with the help of a multi-precision package \cite{bailey}.
Typically, we performed our computations with about 500 significant digits.
In both cases the complex zeros duplicate the result for the thermodynamic
limit. For two flavors, the zeros join into two curves which are separated
by $\sim1/\sqrt n$ for large $n$ and approach the analytical result 
(\ref{mucurve}). Remarkably, half of the zeros coincide almost exactly
with (\ref{mucurve}). The picture is quite different for 
$N_f = 1+1^*$. In this case we observe 
doubly degenerate zeros that approach (\ref{mucurve}), that approach  
(\ref{mucurve}) rotated by 90 degrees, 
and isolated zeros that are located along the diagonals. The fraction of
zeros in each of these three sets remains finite in the thermodynamic limit.
In this figure, the total number in each set is given by, 120, 120 and
80, respectively. Of course, this pattern reflects that for $N_f = 1+ 1^*$
the polynomial is a polynomials in $\mu^4$.
The vertical axis represents the real $\mu$ axis. For $N_f = 2$ we observe
a second order phase transition at $\mu_c \approx 0.52$. This result can
also be obtained by an extrapolation of the distribution of the zeros
of the $N_f = 1+1^*$ partition function. We conclude that this partition
function contains information of the full theory.
For $N_f = 1+ 1^*$ the
density of the zeros decreases linearly with the distance to 
the imaginary axis which  is typical for a
second order phase transition.  
       
\subsection{Phase Diagram}
In the random matrix model one can include a schematic temperature 
dependence 
by the substitution $\mu \rightarrow \mu + i T$. In this case
the $\sigma$- model representation of the partition function is given by
\cite{phase}
\be
Z(m,\mu,T) =\int d\sigma e^{-N{\cal L}(\sigma)}
\ee
where ${\cal L}(\sigma)$ is given by
\be
{\cal L}(\sigma) = {\rm Tr}[\sigma \sigma^\dag
- \frac12\ln\{ [(\sigma+m)(\sigma^\dag+m)- (\mu+iT)^2]\cdot
[(\sigma+m)(\sigma^\dag+m)- (\mu-iT)^2]
\}]\nonumber
\ee
and $\sigma $ is an arbitrary complex $N_f\times N_f $ matrix.
This model was studied for $\mu = 0$ in \cite{JV,Tilo,Stephanov1}. It was
found that it describes a second order phase transition at a nonzero
temperature. Multicritical behavior was studied in
\cite{dammulti,akemann,brezinmulti,janikmulti}.
At zero temperature, a first order phase transition 
\cite{Stephanov} was found at $\mu\ne 0$.

Since, the value order parameter is given by the expectation value $\sigma$, 
this partition function can be interpreted as a Landau-Ginzburg functional.
A similar functional can be derived from a Nambu model \cite{krishna}.
It possesses a tricritical point at $m=0$, $\mu = \mu_3$ and $T= T_3$. 
The critical
exponents at this point are given by the universal mean field critical 
exponents. We emphasize that the upper critical 
dimension at the tricritical point is equal to 3 so that up to logarithmic
corrections, the mean field  critical exponents are exact. 
For a detailed discussion of the phase structure 
of this partition function and the physics of the tricritical point 
\cite{tricritical}
we refer to the talk by M. Stephanov \cite{mish}
in this volume. For other studies of the RMT partition function at
nonzero $\mu$ and $T$ we refer to \cite{JaNo97}. 

\subsection{Random Matrix Triality at Nonzero Chemical Potential}

In this section, we provide evidence that the RMT Dirac operator at $\mu \ne 0$
shows universal behavior as well. In particular, we will consider the chRMT
partition function for all 
three values of $\beta$ and show that the anti-unitary
symmetries of the Dirac operator result in characteristic eigenvalue 
correlations.

Numerical simulations have been performed for all three classes.
A cut along the imaginary axis below a
cloud of eigenvalues was found in instanton liquid
simula\-tions {\cite{Thomas}}  for $N_c =2$ at $\mu \ne 0$
which corresponds to $\beta =1$. In lattice QCD simulations
with staggered fermions for $N_c = 2$ {\cite{baillie}}
a depletion of eigenvalues along the imaginary axis was observed, whereas
for $N_c=3$ the eigenvalue distribution did not show any pronounced 
features \cite{everybody}.

\begin{figure}[!ht]
{\makebox{\epsfig{file=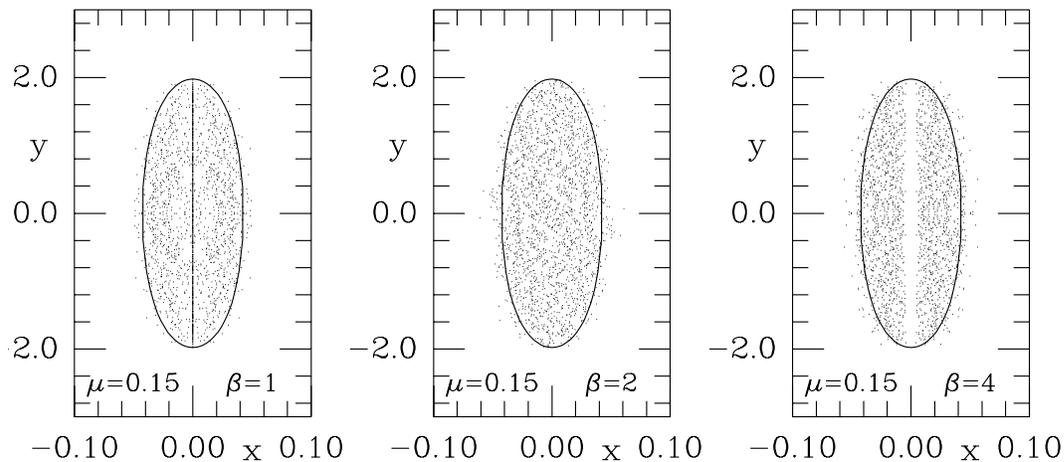,
 width=140mm,angle=0}}}
\vspace*{-1cm}
\caption{
Scatter plot of the real ($x$), and the imaginary
parts ($y$) of the eigenvalues of the random matrix Dirac operator.}
\label{fig6}
\vspace*{-0.5cm}
\end{figure}

In the quenched approximation, the spectral properties
of the random matrix Dirac operator (\ref{Diracmatter})
can easily be studied numerically by  diagonalizing a set of
matrices with probability distribution (\ref{zrandom}).
In Fig. (\ref{fig6}) we show results {\cite{Osbornmu}}
for the eigenvalues of a few
$100\times  100$ matrices for $\mu = 0.15$ (dots). The solid curve represents
the analytical result for the boundary of the domain of eigenvalues
derived in {\cite{Stephanov}} for $\beta =2$. However, 
the method that was used can be extended {\cite{Osbornmu}}
to $\beta = 1$ and $\beta =4$ and with the proper scale factors we find
exactly the same solution.

For  $\beta =1$ and $\beta = 4$ we observe exactly the same structure as in 
the previously mentioned (quenched) simulations.
We find an accumulation of eigenvalues on the imaginary axis for $\beta = 1$
and a depletion of eigenvalues along this axis for $\beta = 4$.
This depletion can be understood as follows. For $\mu = 0$ all eigenvalues
are doubly degenerate. This degeneracy is broken at $\mu\ne 0$ which produces
the observed repulsion between the eigenvalues.

The number of purely imaginary eigenvalues for $\beta = 1$
scales as $\sqrt N$ and is thus not visible in 
a leading order saddle point analysis. 
Such a $\sqrt N$  scaling is typical
for the regime of weak non-hermiticity first identified by Fyodorov
{\it et al.} {\cite{fyodorov}}. 
Using the supersymmetric method of random matrix theory
the $\sqrt N$ dependence was obtained
analytically by Efetov {\cite{Efetovnh}}.
Obviously, more work has to be done in order to
arrive at a complete characterization of
universal features {\cite{fyodorovpoly}} in
the spectrum of nonhermitean matrices.

\section{CONCLUSIONS}
There is strong evidence from lattice QCD simulations and partially quenched
chiral perturbation theory that eigenvalue correlations below a scale
of $F^2/\Sigma \sqrt V$ are given by chRMT. We emphasize that this is an 
exact result. On the other hand, qualitative results have been obtained 
for  a schematic chRMT model for the 
 chiral phase transition at nonzero chemical potential
and temperature.  We have discussed  the mechanism of quenching and
the distribution of the Yang-Lee zeros and found that the phase quenched
partition function contains information of the critical chemical potential
of the full theory. 
\vskip 0.5cm
\noindent
{\bf Acknowledgements}

This work was partially supported by the US DOE grant
DE-FG-88ER40388. Frithjof Karsch and Maria-Paola Lombardo  are thanked 
for organizing this workshop. 
We benefitted from discussions with P. Damgaard, T. Guhr, A. Jackson,
A. Sch\"afer, M. Stephanov, T. Wettig and H. Weidenm\"uller. 
J. Osborn is thanked for a critical
reading of the manuscript.
Finally, I thank my collaborators on whose work this review is based.

\end{document}